\documentstyle[11pt]{article}
\newcommand{\beq}{\begin{equation}}
\newcommand{\eeq}{\end{equation}}
\newcommand{\beqr}{\begin{eqnarray}}
\newcommand{\eeqr}{\end{eqnarray}}

\newcommand{\cosec}{{\rm cosec}}
\newcommand{\sech}{{\rm sech}}
\newcommand{\cosech}{{\rm cosech}}
\newcommand{\sss}{\vspace{.2in}}
\begin{document}
{}~\hfill{UICHEP-TH/95-5}
\sss
\begin{center}
{\Large {\bf Methods for Generating Quasi-Exactly Solvable Potentials}}
\end{center}
\vspace{.8in}
\begin{center}
{\large{\bf Asim Gangopadhyaya$^a$, Avinash Khare$^b$
and Uday P. Sukhatme$^c$}}
\end{center}
\vspace{.8in}

\noindent
a) \hspace*{.2in}
Department of Physics, Loyola University Chicago, Chicago, Illinois 60626\\
b) \hspace*{.2in}
Institute of Physics, Sachivalaya Marg, Bhubaneswar 751005, India\\
c) \hspace*{.2in}
Department of Physics (m/c 273), University of Illinois at Chicago,\\
\hspace*{.4in}845 W. Taylor Street, Chicago, Illinois 60607-7059 \\
\vspace*{1.1in}

\centerline{\large{\bf Abstract}}
 \vspace*{.2in}

We describe three different methods for generating quasi-exactly solvable
potentials, for which a finite number of eigenstates are analytically
known. The three methods are respectively based on (i) a polynomial
ansatz for wave functions; (ii) point canonical transformations; (iii)
supersymmetric quantum mechanics. The methods are rather general and
give considerably richer results than those available in the current
literature.

\newpage

In recent years, many authors have discussed examples of potentials for
which a finite number of eigenstates are explicitly known, but
not the whole spectrum. The
first examples of such quasi-exactly solvable (QES) potentials were
given by Razavy\cite{Razavy80} and others\cite{Singh78,Flessas79}.
Subsequently, using a more general approach, many QES potentials were obtained
by Turbiner, Ushveridze and
Shifman\cite{Turbiner87,Turbiner88,Shifman89} and their connection with
finite dimensional group representations was established
\cite{Turbiner88b,Jatkar89}.
Also, Roy and Varshni\cite{Roy91}
gave an example of the generation of a new QES potential using
supersymmetric quantum mechanics.

In this paper, we develop and discuss three general methods for
generating QES potentials. The first method is a systematic extension of
the approach of Turbiner\cite{Turbiner88}, which has the distinctive
feature of a
polynomial ansatz of the same degree for all analytically known eigenstates,
in addition to a conventional exponential factor for satisfying boundary
conditions. Our extension provides a clearer motivation and
consequently yields many three parameter QES potentials which are more general
than those available in the literature. The method hinges on the choice
of an initial variable $z(x)$ and its use in making a specific ansatz for
wave functions.

The second method consists of making point canonical
transformations (changes of dependent and independent variables)
\cite{Gangopadhyaya94} on
known QES potentials in order to obtain new ones \cite{Lucht93}.
This approach
is motivated from the observation that
all known exactly solvable problems are inter-related via point canonical
transformations.

The third method discussed in this letter applies the techniques
of supersymmetric quantum mechanics\cite{Cooper95} to
calculate the partner potentials of QES
potentials with $(n+1)$ known eigenstates. As expected from the level
degeneracy theorem for unbroken supersymmetry, the supersymmetric partners
are necessarily new
quasi-exactly solvable potentials with $n$ known eigenstates. \sss

\noindent {\bf Method 1: Polynomial Ansatz for Wave Functions:}\sss

This method is a generalization of Turbiner's
approach \cite{Turbiner88,Turbiner88b}
for generating QES potentials.
We want to solve the Schr\"odinger equation ($\hbar=2m=1$)
\beq
-\psi''(x)+\left[ V(x)-E \right]\psi(x)=0~~,  \label{II.1}
\eeq
along with the boundary condition that wave functions for bound states
vanish as
$x$ approaches the end points of the domain.
The initial step in Method 1 for generating
quasi-exactly solvable potentials with $(n+1)$ analytically known
eigenstates is to choose a nonsingular monotonic variable $z(x)$ and write the
wave function $\psi$ in the form
\beq
\psi(x)=p_n(z,a_0, a_1, \ldots, a_{n-1})\;e^{-\int y(x)\,dx},
\label{II.2}
\eeq
where
\beq
p_n(z)=\sum_{j=0}^n a_{j}z^j \;\;\;\;(a_n=1)~~.
\label{II.3}
\eeq
is a polynomial of degree $n$ in the variable $z$. [Note the special
feature that a polynomial of the same degree $n$ is present in all
analytically known wave functions - however, different wave functions
will correspond to different values of $a_j$ and consequently will
have differing numbers of nodes].
The functional form of $y(x)$ remains to be determined. Substituting for
$\psi(x)$ into Eq. (\ref{II.1}) yields
\beq
V(x)-E=\;y^2-y'+{ {p''_n-2yp'_n}\over p_n}~~.
\label{II.4}
\eeq
In order that the potential have no poles in the physical
domain, we require
${p_n''-2y p_n'}$ to have $p_n$ as a factor. Also, in general, one
expects that the energy will be a function of parameters in $y(x)$,
as well as the
coefficients $a_j$ in $p_n$.
However, we do not want these
parameters to show up in the potential, as otherwise a dependence of
the potential on energy would creep in. One way to
prevent this from happening is to
require that ${{p''_n-2y p'_n}\over p_n}$ be a polynomial of degree one.
This condition restricts what the
needed functional form for $y(x)$ ought to be. More specifically, we choose
$y(x)$ by requiring that
the quantity $p''_n-2y p'_n$ be a polynomial of degree $(n+1)$ in $z$.
\beq
p''_n-2yp'_n=\sum_{j=0}^{n+1} b_{j}z^j ~~.
\label{II.5a}
\eeq
This is equivalent to requiring
\beq
(z''-2yz')\sum_{j=1}^n a_j j z^{j-1} + z'^2 \sum_{j=2}^n a_j j(j-1)z^{j-2}
=\sum_{j=0}^{n+1} b_{j}z^j ~~.
\label{II.6}
\eeq
For $n=1$, the $z'^2$ term is absent, and it is sufficient to choose $y(x)$
such that
\beq
z''-2yz'=q_0+q_1z+q_2z^2~~.
\label{II.7}
\eeq
Although the three constants $q_0,q_1,q_2$ in $y(x)$ are arbitrary, their
range of values is usually restricted by the requirement that the
bound state wave function $\psi(x)$ vanish at the end points.
For $n> 1$, the $z'^2$ term in Eq. (\ref{II.6}) does contribute. Therefore,
one needs the additional necessary
constraint that the choice of initial variable $z(x)$ be such that
\beq
z'^2=B_0+B_1z+B_2z^2+B_3z^3~~.
\label{II.8}
\eeq
This above constraint ensures that $p''_n-2yp'_n$ is a polynomial
of order of $(n+1)$. It also constrains the choices allowed
for $z(x)$, which now  must have the form
\beq
\int { {dz} \over \sqrt{B_0+B_1z+B_2z^2+B_3z^3} } = x.
\label{II.9}
\eeq
The variables $z=x^{\pm 2}, \cos\,x, \cosh\,x, e^{\pm x},\sech^2\,x,
\cdots$, etc. used
in references \cite{Razavy80,Turbiner87,Turbiner88,Turbiner88b},
are all special cases of the above Eq. (\ref{II.9}).
Dividing Eq. (\ref{II.5a}) by Eq. (\ref{II.3}) gives
\beq
\frac{p''_n-2yp'_n}{p_n}=b_{n+1}z+(b_n-a_{n-1}b_{n+1})+R_n(z)~~,
\label{II.10}
\eeq
where the remainder polynomial is
\beq
R_n(z)=\sum_{j=0}^{n-1} z^j f_n(j)~~;~~f_n(j) \equiv b_j-a_{j-1}b_{n+1}
-a_j b_n+a_ja_{n-1}b_{n+1}~~.
\label{II.11}
\eeq
Clearly, the requirement of no poles in the potential $V(x)$ implies
\beq
f_n(j)=0~~,~~j=0,1,\ldots,n-1~~.
\label{II.12}
\eeq
These $n$ constraint equations can be solved to determine the constants
$a_0,a_1,\ldots,a_{n-1}$. There are $n$ solutions for any constant
$a_j$, since it satisfies an equation of degree $n$.
The potential $V(x)$ and energies $E$ are given by
\beq
V(x)=y^2-y'+b_{n+1}z(x)~~;~~E=-b_n+a_{n-1}b_{n+1}~~.
\label{II.13}
\eeq
The potential $V(x)$ depends only on the parameters $q_0,q_1,q_2$
appearing in $y$ and $b_{n+1}=nq_2+n(n-1)B_3$. Also, since
$b_n=nq_1+(n-1)q_2a_{n-1}+n(n-1)B_2+(n-1)(n-2)B_3a_{n-1}$,
a useful, alternative form for the energies is
$E=-nq_1-n(n-1)B_2 + a_{n-1}[q_2+2(n-1)B_3]$. Clearly, there are $n$ energy
eigenvalues corresponding to the $n$ values of the constant $a_{n-1}$.

For concreteness, let us fully describe the important special case
corresponding to the value $n=1$. Here, we are constructing QES potentials
with two analytically known eigenstates. The wave functions for these two
states are both of the form given in Eqs. (\ref{II.2}) and (\ref{II.3}).
\beq
\psi(x)=p_1(z,a_0)\;e^{-\int y\,dx},~~p_1(z)=a_0+z(x)~~.
\label{II.14}
\eeq
The quantity $y(x)$ is given by Eq. (\ref{II.7}). For this case,
the constraints given by Eqs. (\ref{II.12}) read
\beq
f_1(0) \equiv q_0-a_0 q_1 +a_0^2 q_2 = 0~~.
\label{II.15}
\eeq
This quadratic equation for $a_0$ has two solutions. The QES potential is
$V(x)=y^2-y'+q_2z(x)$, and it has two analytically known eigenvalues
\beq
E=a_0q_2-q_1=- \frac {q_1}{2} \pm \frac{1}{2} \sqrt {q_1^2-4q_0q_2}~~,
\label{II.16}
\eeq
with corresponding eigenfunctions given by Eq. (\ref{II.14}).

To illustrate the above-described polynomial ansatz method for generating
QES potentials with $(n+1)$ known eigenstates, we give three explicit
examples with different choices of the initial variable $z(x)$. These
examples have been specifically chosen, since the potentials they generate
have not been previously discussed in the literature.

\noindent {\bf Example 1:} $z=x^{\epsilon} , n=1$ [2 known eigenstates].

\noindent  We consider the polynomial $p_1=z+a_0$. Eq. (\ref{II.7})
then gives $$y= { {\epsilon -1}\over {2x} }
-\frac{1}{2\epsilon}
\left(q_2 x^{\epsilon+1}+q_1x+q_0x^{-\epsilon+1}\right)~~.$$
The QES potential with two known eigenstates is
\beqr
V_2(x)&=&{{(\epsilon+1)(\epsilon-1)}\over {4x^2} } -
\frac{1}{2\epsilon}\left[
q_1(\epsilon-2) - 2q_2 (\epsilon+1)x^{\epsilon}
  + 2 q_0 (\epsilon-1)x^{-\epsilon}\right]
\nonumber\\&&
+\frac{1}{4\epsilon^2}\left[
 2 q_0q_1x^{2-\epsilon}
+ 2q_1q_2 x^{2+\epsilon}+
(q_1^2+2 q_0q_2) x^2 + q_0^2 x^{2-2\epsilon}
+q_2^2 x^{2+2\epsilon}\right]~~.
\label{II.18}
\eeqr
The range of the potential is the half-line $0<x<\infty$ [unless
$\epsilon=1$, in which case the range is the full line].
The known energy eigenvalues
are given by Eq. (\ref{II.16}), and the eigenfunctions obtained using
Eq. (\ref{II.14}) are:
\beq
\psi=(x^\epsilon+a_0)
\exp
\left[
{{1-\epsilon}\over {2} } \log x +\frac{1}{2\epsilon}\left(
 {q_2 \over {2+\epsilon} } x^{2+\epsilon}
+ {q_0 \over {2-\epsilon} } x^{2-\epsilon}
+ {1 \over 2} q_1 x^2 \right)
\right]
\;\;{\rm for}\;\epsilon \neq \pm 2.
\label{II.19}
\eeq
The arbitrary constants are constrained by requiring that $\psi$ satisfies
boundary conditions. For instance, for any choice $\epsilon >2$, one needs
$q_0>0$, $q_2<0$. As an explicit example, consider $\epsilon=3, q_0=1,
q_2=-1,q_1=0$. This gives $a_0=\mp 1$,
eigenenergies $E=\pm 1$ and eigenfunctions
$$\frac{(x^3\mp 1)}{x} \exp[-\frac{1}{6}(\frac{x^5}{5} +x^{-1})].$$
Note  that the polynomial $x^3+a_0$
has no zero for the ground state, but one for the first excited state.

\noindent {\bf Example 2:} $z=\cosh x$ , [$(n+1)$ known eigenstates].

\noindent  Since this choice of variable satisfies Eq. (\ref{II.8}),
one can use it
to generate QES potentials with an arbitrary number of known eigenstates.
[The variable $\cosh x$ also appears in Table 4.1 of Ref. \cite{Cooper95},
and corresponds to the exactly solvable generalized P\"oschl-Teller potential
$V(x)=A^2+(B^2+A^2+A)\cosech^2r-B(2A+1)\coth r\,\cosech~r$ with eigenenergies
$E_n=A^2-(A-n)^2].$
First we generate a quasi-exactly solvable potential with two
eigenstates by taking the polynomial
$p_1(x)= \cosh x +a_0$.
This gives the following form for the function $y(x)$:
\beq
y(x)=\alpha \sinh x - \beta \cosech x-\gamma \coth x,
\label{II.20}
\eeq
where $\alpha, \beta, \gamma$ are arbitrary constants.
As before, $a_0$ obeys a quadratic equation
\beq
2\alpha a_0^2+(1+2\gamma)\,a_0-2(\alpha+\beta)=0.
\label{II.21}
\eeq
Its two roots are given by
\beq
a_0={ {-(1+2\gamma)\pm \sqrt{(1+2\gamma)^2+16\alpha
(\alpha+\beta)} } \over {4\alpha}  }.
\label{II.22}
\eeq
These roots give rise to different number of nodes in the polynomial $p_1$ and
hence in the wave functions $\psi$:
$$\psi\propto (\cosh x+a_0) e^{-\alpha \cosh x}
\left( \cosh {x\over 2}\right)^{\gamma-\beta} \left(\sinh
{x\over 2}\right)^{ \gamma+\beta }.$$
The requirement that $\psi$ vanishes at $x=\infty$ and $x=0$ implies that
$\alpha>0$ and $\gamma+\beta>0$.
The potential is
\beq
V_2(x)=\left( \gamma^2 -\gamma +\beta^2 \right) \coth^2 x - \alpha
\left(2 \gamma+3\right)  \cosh x +\alpha^2 \sinh^2x +\beta (2\gamma-1)
\coth x ~\cosech~ x.
\label{II.23}
\eeq
It is independent of $a_0$ and singular at the
origin, where it diverges as
${ {( \gamma+\beta)(\gamma+\beta-1)} \over x^2}$.
The two lowest energy eigenvalues are given by
\beq \label{E}
E=\left( 2  \alpha  \beta + \beta^2 - 3 \gamma-2 \alpha a_0- 1 \right),
\eeq
where, the solutions $a_0$ are given in Eq. (\ref{II.22}).

{}From Eq. (\ref{II.13}), and using $b_{n+1}=nq_2$, we can get a potential with
$n+1$ states
\beq
V_{n+1}=\left( \gamma^2 -\gamma +\beta^2 \right) \coth^2 x - \alpha
\left(2 \gamma+2n+1\right)  \cosh x +\alpha^2 \sinh^2x +\beta (2\gamma-1)
\coth x ~\cosech~ x.
\label{II.24}
\eeq
The energy eigenvalues are given by Eq. (\ref{II.13}), where $a_{n-1}$
satisfies an algebraic equation of degree $n+1$. In particular, for $n=2$,
the equation for $a_1$ ia a cubic:
$$-2 \alpha^2 a_1^3-\alpha (7+6\gamma) a_1^2 + 2(4\alpha^2+4\alpha\beta
-5\gamma -2\gamma^2-3)a_1+ 4(\alpha+2\beta+2\alpha\gamma+2\beta\gamma)=0.$$

Note that a special case of the QES potential of Eq. (\ref{II.24})
corresponding to $\beta=0, \gamma=0,1$ was considered by
Razavy\cite{Razavy80}. The two choices of $\gamma$ give the even and
odd eigenstates respectively. For these situations, the range
of the potential is $-\infty<x<\infty$, since there is no singularity at
$x=0$.
Also, in the limit $\alpha\rightarrow 0$, this model goes to the generalized
P\"oschl-Teller potential.

\noindent {\bf Example 3:} $z=\sech^2x$.

\noindent  This is another example
where we can get a quasi-exactly
solvable potential with more than two known states.
The constants $B_j$ in Eq. (\ref{II.8}) are $B_0=B_1=0, B_2=4, B_3=-4$.
The function $y(x)$ is
$$y=\left(3+\frac{q_2}{2}\right)\cosech\,2x+\left(\frac{q_1}{4}-1 \right)
\coth x +\frac{q_0}{4}\cosh^2x\coth x,$$
and the potential is
given by Eq. (\ref{II.13}) with $b_{n+1}=nq_2-4n(n-1)$.
A special case of this potential was discussed by Turbiner in
Ref. \cite{Turbiner88}. For $n=1$, the
two lowest energy eigenvalues are given by
$E=-q_1 + q_2\,a_0,$
and corresponding eigenfunctions are
$$\psi=\left( \sech^2 x+a_0 \right)
\left( \cosh\,x \right)^{\frac{(q_2+6)}{4}}
\left( \sinh\,x \right)^{ \frac{-(q_2+q_1+q_0+2)}{4} }
\exp\left[- {q_0  \,\cosh\,2x \over 16}  \right].$$
\sss

\noindent {\bf Method 2: Point Canonical Transformations:}\sss

Point canonical transformations (PCT) are known \cite{Gangopadhyaya94} to
transform the  Schr\"odinger equation for a potential $V(x)$
into the Schr\"odinger
equation for a new potential $\tilde V(\xi)$. In this section,
we employ PCT to generate new quasi-exactly
solvable models starting from a known one.
In a point canonical transformation, one changes from the independent
variable $x$ to the variable $\xi$,  where $x=f(\xi)$
and the wave
function is transformed by:
$\psi(\alpha_i; x) \longrightarrow\left({ {df}\over {d\xi} }
\right)^{1/2} \tilde{\psi}  \left( \alpha_i; \xi \right)$.
The new potential and energy (denoted with tilde on the top) are then
given by:
\beq
\tilde{V}-\tilde{E} =
f'^2 \left[ V(\alpha_i; f(\xi)) - E(\alpha_i) \right]
 + {1 \over 2}
\left( {3f''^2 \over 2f'^2} - {f''' \over f'}  \right)
\label{II.25}
\eeq
The constant term on the right hand side represents the energy of the new
potential. It should be emphasized that the above-described
technique is quite general. We will illustrate it with a specific example.

\noindent{\bf Example 4:}

\noindent  As a starting point, we choose the new QES potential
$V_2(x)$ obtained
in Eq. (\ref{II.23}) of Example 2.
The change of variables $x=i\xi$ converts hypergeometric functions into
trigonometric functions. The new potential for $0<\xi<\pi$ is:
\beq
\tilde{V}(\xi)=\left( \gamma^2 -\gamma +\beta^2 \right) \cot^2 \xi + \alpha
\left(2 \gamma+3\right)  \cos \xi +\alpha^2 \sin^2\xi +\beta (2\gamma-1)
\cot \xi \cosec \xi.
\eeq
The two known eigenvalues of this potential $\tilde{E}$ are given by $-E$,
with $E$ given by Eq. (\ref{E}).
The corresponding wave functions are given by
$$\psi\propto (\cos \xi + a_0) e^{-\alpha \cos \xi}
\left( \cos{\xi \over 2}\right)^{\gamma-\beta} \left(\sin
{\xi\over 2}\right)^{ \gamma+\beta }~.$$
The constraints $\gamma-\beta>0, \gamma+\beta>0$ ensure that $\psi$ vanishes
at the end points.

Alternatively, if we make the change of variables
$x=\coth^{-1} [i \sinh\,\xi]$, then
the new potential is
\beqr
\tilde{V}&=&\left(3\alpha - \beta+ 2\alpha\gamma + 2\beta
\gamma\right)\tanh\xi -
\left(2\alpha^2 -\beta^2+ E+\gamma-
\gamma^2\right)\tanh^2\xi
\nonumber\\&&
-\left(3\alpha+2\alpha\gamma\right)
\tanh^3\xi + \alpha^2\tanh^4\xi.
\eeqr
This is a well defined QES potential over the range
$-\infty<\xi<\infty$. The two lowest eigenvalues are
$\tilde E = -\alpha^2-E$ , where $E$ is given by Eq. (\ref{E})
and the eigenfunctions are
\beq
\psi \propto (\tanh z+a_0) e^{-\alpha \tanh z} (\tanh z)^{\gamma+\beta}
(\sech z)^{\gamma-\beta-\frac{1}{2}}~.
\eeq
\sss

\noindent {\bf Method 3: Supersymmetric Quantum Mechanics:}\sss

If one has a potential $V_-(x)$ whose lowest eigenstate has energy
$E_0^{(-)}=0$ and wave function $\psi_0(x)$, then its supersymmetric
partner potential $V_+(x)$ is given by \cite{Cooper95}
\beq \label{II.29}
V_+(x)=V_-(x)-2 \frac{d}{dx}
\left( \frac{\psi_0'}{\psi_0} \right)~~.
\eeq
The degeneracy theorem for unbroken supersymmetry states that
$V_+$ and $V_-$ have the same energy levels (except for the zero energy
ground state) and simply related eigenfunctions:
\beq \label{II.30}
E_{k-1}^{(+)}=E_k^{(-)}~;~\psi_{k-1}^{(+)}={E_k^{(-)}}^{-\frac{1}{2}}
\left( \frac{d}{dx}-\frac{\psi_0'}{\psi_0}  \right)~
\psi_k^{(-)}~;~k=1,2,\ldots
\eeq

Clearly, if one applies the above results to a QES potential with $(n+1)$
known eigenstates, one gets a new QES potential with $n$ known eigenstates.
For the simplest case of $n=1$ and $V_-(x)=x^6-7x^2+2 \sqrt{2}$, this
procedure was discussed in Ref. \cite{Roy91}.

As a general illustration, we want to apply the supersymmetry approach to a
QES potential obtained by Method 1 using a polynomial ansatz for the
known wave functions. Since Eq. (\ref{II.2}) gives
$$\psi_0 \propto p_0 e^{-\int y(x)~dx},$$
the superpotential $-\psi_0'/\psi_0$ appearing in
Eqs. (\ref{II.29}) and (\ref{II.30}) is just
$y(x)-p_0'/p_0$, where the parameters $a_j$ appearing in the polynomial
$p_0$ correspond to the ground state. The supersymmetric partner potential is
\beq \label{II.31}
V_+(x)=V_-(x)+2y'(x)-\frac{2(p_0p_0''-p_0'^2)}{p_0^2}
\eeq
and the lowest $n$ eigenfunctions are
\beq
\psi_{k-1}^{(+)} \propto \left(   \frac{p_k'}{p_k} - \frac{p_0'}{p_0}
\right) p_k~e^{-\int y(x) dx }~,
{}~k=1,2,\ldots,n~.
\eeq
where $p_k$ is the polynomial $p(z)$ with parameters $a_j$ corresponding
to the $k^{\rm th}$ state of $V_-(x)$.
Note that the same procedure of generating a supersymmetric partner
potential can again be repeated by starting from the potential $V_+(x)$.
Also, the standard methods of supersymmetric quantum mechanics
\cite{Cooper95} can be used to
generate multi-parameter isospectral potential families
of $V_-(x)$, which are of course new QES potentials.

\sss

A.G. and A.K. acknowledge the hospitality of the UIC
Department of Physics where
part of this work was done. Partial financial support from the U.S. Department
of Energy is gratefully acknowledged.

\sss

\end{document}